\documentstyle[prl,aps,epsfig]{revtex}

\newcommand{\cms}{cm$\;\mbox{s}^{-1}\;$~}
\newcommand{\kms}{km$\;\mbox{s}^{-1}\;$~}  
\newcommand{\cmsgm}{cm$^2\;\mbox{gm}^{-1}\;$~}

\begin{document}

\draft  
\tighten
\twocolumn[\hsize\textwidth\columnwidth\hsize\csname@twocolumnfalse%
\endcsname

\title{Gravothermal Collapse of Self-Interacting Dark Matter Halos\\ 
and the Origin of Massive Black Holes}

\author{Shmuel Balberg$^1$ and Stuart L.~Shapiro$^2$}

\address{Physics Department, University of Illinois at Urbana-Champaign, 
1110 W.~Green st., Urbana, IL 61801}
\address{$^1$also Racah Institute of Physics, The Hebrew University, 
Jerusalem 91904, Israel}
\address{$^2$also Department of Astronomy and NCSA, 
University of Illinois at Urbana-Champaign, 
Urbana, IL 61801}
\maketitle 

\begin{abstract}
Black hole formation is an inevitable consequence of relativistic core 
collapse following the gravothermal catastrophe in self-interacting dark 
matter (SIDM) halos. Very massive SIDM halos form supermassive black holes 
(SMBHs) $\gtrsim 10^6\;M_\odot$ directly. Smaller halos believed to form by 
redshift $z=5$ produce seed black holes of $10^2-10^3\;M_\odot$ 
which can merge and/or accrete to reach the observational SMBH range. 
This scenario for SMBH formation requires no baryons, no prior star 
formation and no other black hole seed mechanism.
\end{abstract}

]

\narrowtext

Black holes in the centers of galaxies are inferred to have masses of
$10^6-10^9\;M_\odot$, and to date dozens of candidates have been discovered 
\cite{SMBHs}. Black holes in this mass range 
are the likely power sources in quasars and active galactic nuclei 
\cite{ZelSal64,Rees84,Quasars}. Recent observational data have identified a 
strong correlation between the inferred black hole mass and the stellar 
velocity dispersion in the host galaxy bulge \cite{sigcor}; a correlation has 
also been found between the black hole mass and the mass of the bulge 
\cite{bulgecor}. Both correlations suggest that formation and evolution of the 
central black hole and the bulge of the host galaxy may be closely related.

The origin of these supermassive black holes (SMBHs) is uncertain. 
Proposed scenarios include the collapse of a supermassive star 
\cite{SMS}, possibly built up by collisions and 
mergers of ordinary stars \cite{QuinShap90}. Another scenario involves the 
collapse of the core of a dense relativistic cluster (e.g.~neutron stars or 
stellar mass black holes) following gravothermal evolution.  The cluster core 
evaporates mass to an extended halo on a 
gravitational scattering {\it relaxation} timescale,
while the core density and velocity dispersion grow (the ``gravothermal 
catastrophe''; \cite{LBW68,Spitzer87}). Zel'dovich and Podurets originally 
conjectured \cite{ZelPod65} and Shapiro and Teukolsky 
\cite{ShapTeu} subsequently demonstrated that a nearly 
collisionless gas in virial equilibrium like a star cluster experiences a 
radial instability to collapse on a {\it dynamical} timescale when its core 
becomes sufficiently relativistic. As the instability sets in, the core and 
its immediate surroundings undergo catastrophic collapse to a black hole 
and the ambient halo settles into a new dynamical equilibrium state around 
the central hole. 

Here we report that the formation of a black hole in the center of a galactic 
halo is a natural and inevitable consequence of the gravothermal evolution of 
self-interacting dark matter (SIDM) halos. The possibility that dark matter 
particles are ``self-interacting'' has been revived recently 
\cite{SSSIDM}. Studies of SIDM via N-body simulations 
\cite{Yoshidaal00,Daveal01} and via a gravothermal approach \cite{SIDMGT} have 
confirmed that SIDM halos are more consistent with observations: they exhibit 
a flat density core, rather than a cuspy one that arises for cold dark matter 
\cite{NFW97,Mooreal99}. An isolated SIDM halo evolves gravothermally, since 
the thermal relaxation timescale due to collisions is shorter than the typical 
ages of cosmological halos. Unlike a star cluster, a SIDM halo retains an 
appreciable core mass as it evolves towards the relativistic 
instability \cite{SIDMGT}. 

\section{Gravothermal Evolution and Collapse of a SIDM halo}

The structure of 
a relaxed SIDM halo can be defined by the central mass density $\rho_{\rm c}$, 
the (one-dimensional) velocity dispersion $v_{\rm c}$, and the total halo mass 
$M_{\rm tot}$. For a given cosmology, the background density, $\bar{\rho}$, 
is a function only of redshift. We assume that the central 
density contrast $x\equiv \rho_{\rm c}(t=0)/\bar{\rho}$ at halo virialization 
is independent of redshift and  halo mass, which is consistent with the 
simulations of Dav$\grave{\mbox{e}}$ et al.~\cite{Daveal01} and recent work on 
truncated isothermal spheres by Shapiro, Iliev and Raga \cite{IlShap1}. The 
independent parameters of a SIDM halo are then $M_{\rm tot},\; x$ and $z_0$, 
the redshift of virialization. Define mass and radius 
scales, $M_0$ and $R_0$ respectively, where $M_0=4\pi R_0^3 \rho_{\rm c}$ and 
$v_{\rm c}^2=G M_0/R_0$, and also the ratio $g\equiv M_{\rm tot}/M_0$, which 
is a unique function of $x$. The central velocity satisfies
\begin{equation}\label{eq:v_c}
v_{\rm c}^2(t=0)=(4\pi)^{1/3} G (M_{\rm tot}/g(x))^{2/3} 
(x \bar{\rho}(z_0))^{1/3}\;.
\end{equation}

{\it Evolutionary Stages.} The properties in the halo depend 
critically on the ratio of the collisional mean-free-path, $\lambda$, 
to the gravitational scale height, $H$:
\begin{equation}\label{eq:lambda-H}
\lambda=(\rho \sigma)^{-1},\;\;\; 
H=\left(\frac{v^2}{4\pi G \rho}\right)^{1/2}\;,
\end{equation}
where $\rho$ is the local mass density, $v$ is the local one dimensional 
velocity dispersion, and $\sigma$ is the SIDM cross section per unit mass. In 
the long mean-free-path (lmfp) region $\lambda\gg H$, particles 
perform several orbits between collisions; in the short mean-free-path 
(smfp) region $\lambda\ll H$, particle motion is severely 
restrained, and heat transfer proceeds through diffusion, as in a fluid. 
Simulations show that a halo core must form 
in the lmfp regime \cite{SSSIDM}; otherwise the particles  
generate strong shocks, whereby entropy loss to the halo destroys the 
central core structure \cite{Mooreal00}. 

The requirement that at formation the core satisfies 
$(\lambda/H)_{\rm c}\geq 1$ yields an {\it upper} limit for the halo mass:  
\begin{equation}\label{eq:needlamH}
M_{\rm tot}\leq 4\pi (\bar{\rho}(z_0)x)^{-2} g(x) \sigma^{-3}\;.
\end{equation}

\begin{figure}[hbt]
\centerline{\epsfig{file=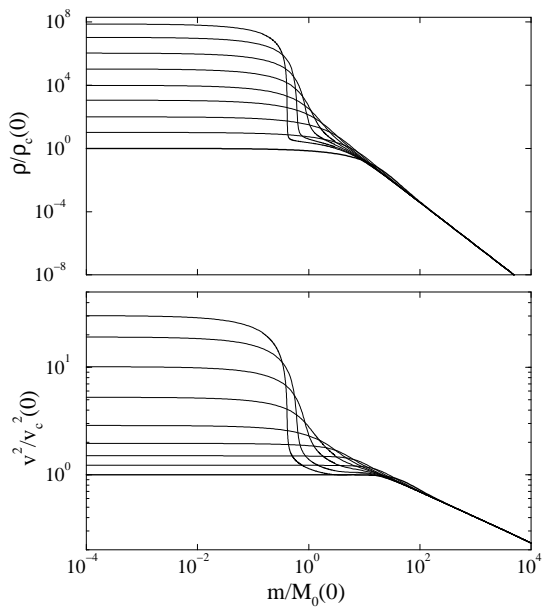,width=0.4\textwidth,angle=0}}
\caption{Snapshots of the (a) density and (b) velocity dispersion profiles
of a SIDM halo with $\sigma=0.67 4\pi R_0^2/M_0$ (for $a=2.26$) at selected
times during gravothermal evolution [16]. The thick line denotes the profile 
at $t=0$. Subsequent profiles correspond to
$t/t_{\rm r}(0)=$264.0,~287,~289.5,~289.90,~290.12,~290.30,~290.49,
and 290.63.}
\label{fig:GTevolve}
\end{figure}

{\it Black Hole Formation} Gravothermal evolution of a SIDM halo is 
shown in Figure~\ref{fig:GTevolve} \cite{SIDMGT}. While the extended halo 
($\rho \propto r^{-2.19}$) remains in the lmfp limit, 
the core becomes increasingly dense. The core eventually bifurcates into a 
smfp inner part which is fluid-like $(\lambda\ll H)$, and a lmfp outer part 
which is nearly static. The transition region corresponds to 
$\lambda/H\sim 1$. It is the inner core and its immediate surroundings which 
collapse to a black hole following the onset of the relativistic 
instability, leaving the outer core and extended halo in dynamical equilibrium
about the central hole \cite{ShapTeu}. 

Evaporation of mass due to collisions from a lmfp core into the halo is rapid 
with $d\log M_{\rm c}/d\log(v_{\rm c}^2)\approx-4.27$  \cite{SIDMGT}. But once 
the core becomes very dense, with central $\lambda/H\gtrsim 100$, 
evaporation is limited to a surface effect, which is much less efficient.
We find \cite{SIDMGT} that if the core of a SIDM halo forms with 
$\lambda/H\geq 1$, its mass decreases by about one order of magnitude prior to 
reaching the smfp limit (while the central velocity hardly changes), 
\hspace{-0.4cm}after which the core mass decreases according to 
$d\log M_{\rm c}/d\log(v_{\rm c}^2)\approx -0.85$. This estimate gives
the inner core mass $M_{\rm coll}$ at the onset of relativistic 
instability ($v_{\rm c} \approx c/3$):
\begin{equation}\label{eq:M_coll} 
\log_{10}\left(\frac{M_{\rm coll}}{M_{\rm c}(0)}\right)
\approx -1+0.85\log_{10}\left(\frac{v_{\rm c}^2(0)}
{10^{10} ~{\rm km^2} ~{\rm s^{-2}}}\right)\;.
\end{equation}

{\it Core Lifetime.} The lifetime of the core until 
collapse is \cite{SIDMGT} $t_{\rm coll}\approx 290 t_{\rm r}(0)$, where 
$t_{\rm r}(0)$ is the collision relaxation timescale in the core at formation, 
\begin{equation}\label{eq:t_r}
t_{\rm r}
=\frac{1}{a}\left(\rho_{\rm c}(t=0) v_{\rm c}(t=0) \sigma\right)^{-1}\;
\end{equation}
and $a\simeq 2.26$ for purely elastic, hard-sphere 
collisions and a Maxwell-Boltzmann velocity distribution. Requiring that the 
core of a halo formed at redshift $z_0$ collapses by redshift $z_1$ yields
a {\it lower} limit on the mass of the SIDM halo:
\begin{equation}\label{eq:needtr}
M_{\rm tot} \geq \left(\frac{290}{a}\right)^3(4\pi G^3)^{-1/2}
    \frac{(\bar{\rho}(z)x)^{-7/2}g(x) \sigma^{-3}}
   {\left[t(z_1)-t(z_0)\right]^3}\;.
\end{equation}

\section{A Direct Scenario for SMBH Formation}

Since the core virializes with 
nonrelativistic velocities, $v_{\rm c}\leq 10^2-10^3\;$\kms, the core mass at 
collapse will be several orders of magnitude smaller than 
$M_{\rm c}(0) \approx 10^{-2} M_{\rm tot}$ according to Eq. ~(\ref{eq:M_coll}).
Consequently, gravothermal 
evolution directly culminating in a SMBH with a mass above $10^6\;M_\odot$ 
requires that the total mass of the progenitor SIDM halo exceed 
$10^{12}\;M_\odot$. Equations~(\ref{eq:needlamH}) and (\ref{eq:needtr}) 
determine the range of halo masses formed at redshift $z_0$ that undergo 
core collapse by redshift $z_1$. These masses increase for smaller 
values of the ratio $x$, i.e., for smaller central density contrasts at halo 
formation. 

\begin{figure}[hbt]
\centerline{\epsfig{file=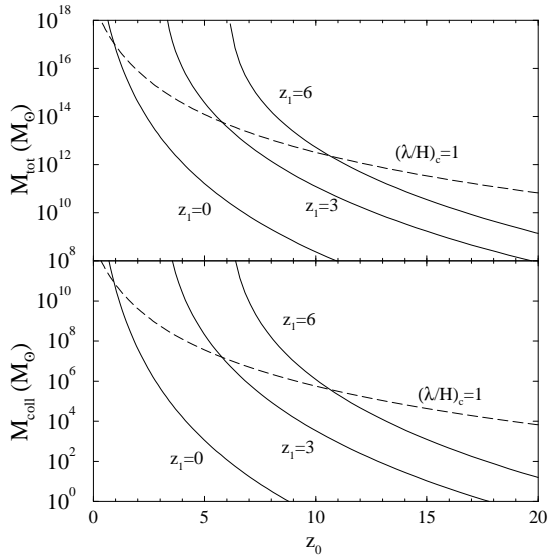,width=0.4\textwidth,angle=0}}
\caption{Masses of SIDM halos that can undergo core collapse by redshifts 
$z_1=6,\;3$ and 0 (upper frame), and the corresponding core mass at collapse
(lower frame), as functions of formation redshift $z_0$. Dashed curve 
corresponds to the upper limit (Eq.~[\ref{eq:needlamH}]) and solid 
curves to the lower limits (Eq.~[\ref{eq:needtr}]) as functions of $z_0$ and 
$z_1$. In this model the density contrast 
$x\equiv \rho_{\rm c}(t=0)/\bar{\rho}=1.8\times 10^4$.}
\label{fig:MsIS}
\end{figure}

We examine this scenario by setting $x$ to the relatively small value of 
$x=1.8\times 10^4$, which leads to $g\approx 206$. This is the value derived 
for a truncated isothermal sphere of dark matter 
\cite{IlShap1} for an Einstein de-Sitter universe, and also applies 
for all but very small redshifts for a flat universe with a finite 
cosmological constant \cite{IlShap2}. In Figure~\ref{fig:MsIS} we show the 
range of halo masses which experience core collapse (upper frame) as a 
function of formation redshift $z_0$. Since the lower limit depends also 
on the desired redshift at collapse, $z_1$, we show three variations: 
$z_1=6,3$ and 0. The lower frame shows the corresponding core ($\approx$ black 
hole) masses at collapse, using 
Eqs.~(\ref{eq:v_c}) and (\ref{eq:M_coll}). The 
cosmological model is $\Omega_{\rm m}=0.3,\;\Omega_\Lambda=0.7$ and $h=0.65$, 
and we set $\sigma=5\;$\cmsgm \cite{Daveal01}. 

We find that SMBHs of $10^6-10^7\;M_\odot$ can form in massive halos which 
experience core collapse at intermediate and low redshifts. At redshift 
$z_1=6$, our model limits the initial black hole mass to 
$M \approx 5\times 10^5 M_\odot$. Subsequent accretion onto the black hole 
must be invoked to reach masses of $10^9\;M_\odot$ \cite{HaiLoeb01}, 
especially to reconcile with recent observations of high red shift quasars
\cite{QUASz6}.
Very young halos are practically excluded as candidates for reaching core 
collapse by redshift $z=0$: halos formed later than $z\approx 1.5$ will not 
have reached core collapse, regardless of mass, and only cores of extremely 
massive halos ($M_{\rm tot}\geq 10^{14}\;M_\odot$) could have collapsed if 
formed after $z=3$.

\section{A Bottom-Up Scenario for SMBH Formation}

\begin{figure}[hbt]
\centerline{\epsfig{file=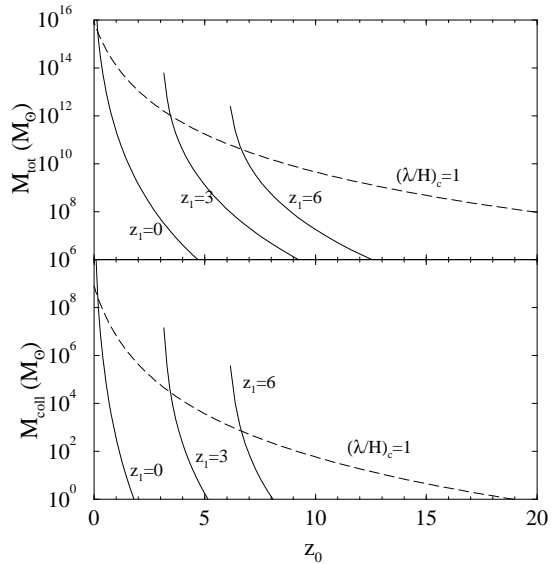,width=0.4\textwidth,angle=0}}
\caption{Same as Figure~\ref{fig:MsIS}, except that 
$x\equiv \rho_{\rm c}(t=0)/\bar{\rho}=10^6$.}
\label{fig:MsMIL}
\end{figure}

Large values of $M_{\rm coll}$ require very massive halos in the 
direct scenario. Virialization of massive halos, especially at high 
redshift, is not favored in the currently accepted picture of structure 
formation \cite{PS,LacCol}. The smaller mass halos formed preferentially in 
current theory can undergo core collapse in a Hubble time provided the 
fiducial value of $x$ is larger 
(see Eqs.~[\ref{eq:needlamH},\ref{eq:needtr}]). In Figure~\ref{fig:MsMIL} we 
show similar results to those of Fig.~\ref{fig:MsIS}, except that now $x=10^6$ 
(and $g\approx 902$). This value appears to be in better agreement with the 
numerical simulations of ref.~\cite{Daveal01}, at least for low redshifts. 

The relations $M_{\rm min}\propto x^{-7/2}$ and $M_{\rm max}\propto x^{-2}$ 
show why core collapse shifts to relatively low mass halos for $x=10^6$. For 
this $x$ any halo which virializes at redshift $z\geq 5$ will have undergone 
core collapse. However, young halos formed at low redshifts will still be safe 
from core collapse, as we found also for $x=1.8\times 10^4$. In this scenario, 
core collapse even by redshift $z=6$ is quite possible, for halos with masses 
in the range $10^7-10^{10}\;M_\odot$ virialized at redshifts $z=7-10$. The 
masses of the collapsed cores are small, and for halo masses considered 
realistic in the Press-Schechter formalism \cite{PS}, $
M_{\rm coll}\leq 10^2-10^3 \;M_\odot$. In this case gravothermal core 
collapse does not lead directly to the formation of SMBHs found in the centers 
of galaxies, but the resulting intermediate mass black holes can serve as 
seeds for SMBH build-up through mergers \cite{Rees84,HaiLoeb01}, or accretion.

\section{Discussion}

Gravothermal core collapse in a SIDM halo triggers the formation of a SMBH at 
its center. The black hole forms when the inner core becomes relativistic and 
dynamically unstable. The mass of the collapsing core will be 
$10^{-8}-10^{-6}$ times the total mass of the halo. Forming SMBHs by core 
collapse in SIDM halos requires no baryons, no prior epoch of star formation 
and no other black hole seed mechanism.

For massive ($>10^{13}\;M_\odot$) halos which virialize at modest 
overdensities, SMBHs with masses of $10^6-10^7\;M_\odot$ form directly 
through gravothermal collapse. An alternative scenario where the smaller SIDM 
halos preferred in current cosmological models reach core collapse in a 
Hubble time is also possible and arises if the typical overdensity of the 
halos at virialization is large. In this case any halo which forms prior to 
redshift $z=5$ produces a black hole by $z=0$, and black hole formation 
at high redshift is also possible. However, these halos give rise only 
to low and intermediate mass black holes, $\leq 10^2-10^3\;M_\odot$. These 
might be seeds of SMBHs through multiple halo mergers \cite{HaiLoeb01}. Even 
if only a few percent of the halos achieve core collapse by $z=5$, this 
initial black hole population is sufficient to generate the observed 
SMBH spectrum eventually \cite{MHN01}. Coalescence of black holes in halo 
mergers is a likely source of gravitational waves, potentially detectable  
with the Laser Interferometer Gravitational Wave Observatory (LIGO) as 
well as the planned Laser Inferometer Space Antenna (LISA). 
If most of the SMBH population arises through 
multiple mergers of lower mass seed black holes, the rate may be as large as 
several per year \cite{MHN01}.

The newly formed black hole may grow through accretion of SIDM from the halo. 
The main reservoir for accretion is the outer core, which is characterized by 
$\lambda/H>1$, and includes a mass $\gtrsim 10^{-3}M_{\rm tot}$. If 
the ambient SIDM halo eventually dominates the spherical bulge of galaxies, 
then the accretion of the entire outer core would provide a natural means of 
producing the ratio $M_{\rm BH}/M_{\rm bulge}$ observed in nearby galaxies. It 
would also explain the origin of the most massive ($>10^9\;M_\odot$) central 
black holes. The possibility that SIDM accretion onto seed black holes is 
the origin of SMBHs was also proposed in \cite{OstHen001}, but there it is 
assumed that the seeds arise from supernovae explosions in massive 
stars rather than the gravothermal scenario discussed here. Their 
quantitative results depend crucially on an assumed singular power-law density 
profile of the SIDM, which has yet to be verified self-consistently.

Young halos, especially low mass ones, could not have reached core 
collapse by $z=0$. This allows SIDM halos to explain the flat density cores 
observed in some 
dwarf and low-surface-brightness galaxies, where the 
inferred values of central density and velocity dispersion are  
$\rho_{\rm c}\approx 0.02\;M_\odot\;\mbox{pc}^{-3}$ and 
$v_{\rm c}\lesssim 10^7\;$\cms, yielding a gravothermal core 
lifetime that greatly exceeds the Hubble time. 

To identify the main route for SMBH formation, it would be useful to determine 
the SMBH population as a function of redshift. More data about the presence or 
absence of central black holes in dwarf galaxies would also be useful.

We thank Y.~Birnboim and P.~R.~Shapiro for useful discussions.
This work was supported in part by 
NSF Grant PHY-0090310 and NASA Grants NAG5-8418 and NAG5-10781 at the 
University of Illinois at Urbana-Champaign.

\end{document}